\documentclass[aps,prb,reprint,amssymb,superscriptaddress,showpacs]{revtex4-1}%
\usepackage{graphicx}
\usepackage{epstopdf}
\usepackage{dcolumn}
\usepackage{bm}
\usepackage{color}
\usepackage{amsmath}
\usepackage{amsfonts}
\usepackage{amssymb}%
\def\B {\color{blue}}
\providecommand{\U}[1]{\protect\rule{.1in}{.1in}}
\begin{document}
\title[Coupling of magnetic order]{Coupling of magnetic order to
planar Bi electrons in the anisotropic Dirac metals
\emph{A}MnBi$_{2}$ (\emph{A} = Sr, Ca)}
\author{Y. F. Guo}
\affiliation{Department of Physics, University of Oxford, Clarendon Laboratory, Parks Road,
Oxford, OX1 3PU, United Kingdom}
\author{A. J. Princep}
\affiliation{Department of Physics, University of Oxford, Clarendon Laboratory, Parks Road,
Oxford, OX1 3PU, United Kingdom}
\author{X. Zhang}
\affiliation{Beijing National Laboratory for Condensed Matter Physics, Institute of
Physics, Chinese Academy of Sciences, Beijing 100190, China}
\author{P. Manuel}
\affiliation{ISIS Facility, Rutherford Appleton Laboratory, Chilton, Didcot, OX11 0QX,
United Kingdom}
\author{D. Khalyavin}
\affiliation{ISIS Facility, Rutherford Appleton Laboratory, Chilton, Didcot, OX11 0QX,
United Kingdom}
\author{I. I. Mazin}
\affiliation{Naval Research Laboratory, code 6390, 4555 Overlook Avenue SW,
Washington DC 20375, USA}
\author{Y. G. Shi}
\email[]{ygshi@aphy.iphy.ac.cn}
\affiliation{Beijing National Laboratory for Condensed Matter Physics, Institute of
Physics, Chinese Academy of Sciences, Beijing 100190, China}
\author{A. T. Boothroyd}
\email[]{a.boothroyd@physics.ox.ac.uk}
\affiliation{Department of Physics, University of Oxford, Clarendon Laboratory, Parks Road,
Oxford, OX1 3PU, United Kingdom}
\date{\today}

\begin{abstract}
We report powder and single crystal neutron diffraction measurements of the
magnetic order in $A$MnBi$_{2}$ ($A$ = Sr and Ca), two layered manganese
pnictides with anisotropic Dirac fermions on a Bi square net. Both materials
are found to order at $T_{\rm N} \approx 300$\,K in $\mathbf{k} = 0$ antiferromagnetic structures, with
ordered Mn moments at $T=10$\,K of approximately 3.8\,$\mu_{\mathrm{B}}$
aligned along the $c$ axis. The magnetic structures are N\'{e}el-type within
the Mn--Bi layers but the inter-layer ordering is different, being
antiferromagnetic in SrMnBi$_{2}$ and ferromagnetic in CaMnBi$_{2}$. This
allows a mean-field coupling of the magnetic order to Bi electrons in
CaMnBi$_{2}$ but not in SrMnBi$_{2}$. We find clear evidence that magnetic
order influences electrical transport. First principles calculations
explain the experimental observations and suggest that the mechanism for
different inter-layer ordering in the two compounds is the competition between
the anteiferromagnetic superexchange and ferromagnetic double exchange carried
by itinerant Bi electrons.
\end{abstract}

\maketitle










\section{Introduction}

Dirac materials are a new class of quasi-two-dimensional electron systems
whose properties are dominated by quasiparticles (Dirac fermions) whose energy
disperses linearly with momentum. In isotropic Dirac materials, such as
graphene, $d$-wave superconductors and topological insulators, the crossing of
linearly dispersing bands at the Dirac point forms a Dirac cone. This,
together with the well defined helicity of the states near the Dirac point, is
responsible for the interesting and unusual behavior observed in Dirac
materials, especially their transport properties in an external magnetic field
\cite{CastroNeto-RMP-2009,Cayssol-CRPhys-2013}.

Recently, the layered manganese pnictides $A$MnBi$_{2}$, with $A$ = Sr and Ca,
were reported to exhibit anomalous metallic behavior consistent with a highly
anisotropic Dirac dispersion and a sizable gap at the Dirac point due to
spin--orbit coupling \cite{Park-PRL-2011,Wang-PRBR-2012}. These compounds are
structurally similar to the iron-based superconductors
\cite{Johnston-AdvPhys-2010,Stewart-RMP-2011} and to novel dilute magnetic
 semiconductors\cite{DMS}. They contain a layer of Mn-Bi composed
of edge-sharing tetrahedra, and a Bi square net, separated by a layer of $A$
atoms. Depending on $A,$ the Mn-Bi layers can be stacked with or without a translation through $(0.5,0.5,0)$, forming correspondingly the $I4/mmm$ or $P4/nmm$ symmetry
groups. First principles density function theory (DFT) band calculations\cite{Park-PRL-2011,Wang-PRB-2011,Wang-PRBR-2012,Lee-PRB-2013}
indicate that Mn is divalent, has five $d$ electrons that are fully spin-polarized,
and that the Dirac states, as well as other bands crossing the Fermi level, arise
from the crossing of folded Bi $6p_{x,y}$ bands in the doubled Bi square net
of $A$MnBi$_{2}$. Interestingly, the Dirac cones are highly anisotropic in the
$xy$ plane, due to weak hybridization with $A$ site $d_{xy,yz}$ orbitals. A substantial amount of
experimental evidence for anisotropic Dirac fermions {\B i}n the Bi layer exists
from measurements of magnetization, magnetotransport, angle-resolved
photoemission spectra (ARPES) and magnetothermopower
\cite{Park-PRL-2011,Wang-PRBR-2011,Wang-PRBR-2012,He-APL-2012,Wang-PRB-2011,Wang-APL-2012,Feng-arXiv-2013}. Other bands predicted by DFT are also seen in ARPES.

A further interesting feature of the $A$MnBi$_{2}$ Dirac materials is the
presence of antiferromagnetic (AFM) order, indicated by anomalies in the
susceptibility at temperatures just below room temperature (SrMnBi$_{2}$:
Refs.~\onlinecite{Park-PRL-2011,Wang-PRB-2011}; CaMnBi$_{2}$: Refs.~\onlinecite{Wang-PRBR-2012,He-APL-2012}). Magnetism is potentially important in Dirac materials because long-range magnetic order could couple to the Dirac fermions and influence electrical transport. DFT
calculations for $A$MnBi$_{2}$\cite{Park-PRL-2011,Wang-PRBR-2012,Wang-PRB-2011,Lee-PRB-2013}
indicate that the ordered moment is carried by the Mn atoms and is
approximately 4\thinspace$\mu_{\mathrm{B}}$ in magnitude,
hybridization-reduced from the value of 5\thinspace$\mu_{\mathrm{B}}$ expected
for localized Mn$^{2+}$ ($3d^{5}, S = 5/2$). Strong in-plane superexchange
leads to N\'{e}el-type antiferromagnetism in the $ab$ plane, and the sense
of the anisotropy in the susceptibility observed in the AFM phase suggests
that the moments point parallel to the $c$ axis. There are no
predictions for the propagation of the magnetic structure along the $c$ axis.
The inter-layer magnetic coupling should be weak, and its sign is hard to predict from
general considerations.

Here we report a neutron diffraction and electrical transport study in which
we establish the three-dimensional magnetic structures of SrMnBi$_{2}$ and
CaMnBi$_{2}$ and observe an anomaly at the AFM transition in the resistivity
of CaMnBi$_{2}$, but not SrMnBi$_{2}$. We find N\'{e}el-type AFM order within
the Mn--Bi layers with a reduced moment, consistent with previous DFT
calculations, but we find two different ordering sequences in the out-of-plane
direction: antiferromagnetic in SrMnBi$_{2}$ and ferromagnetic in CaMnBi$_{2}%
$. This means that coupling between the Mn magnetic order and the Bi square
net (responsible for the electronic transport) is allowed at the mean field
level in CaMnBi$_{2}$ but not in SrMnBi$_{2}$, consistent with the behavior of
the resistivity. Our first principles DFT calculations reproduce the observed inter-layer magnetic order and suggest a microscopic explanation for the differences in the magnetic order and behavior of the resistivity between the two materials.

\section{Experimental and Computational Details}

Polycrystalline samples of CaMnBi$_{2}$ and SrMnBi$_{2}$ were prepared by
solid-state reaction. Stoichiometric amounts of Mn (99.9\%), Bi (99.99\%), and
either Ca (99.99\%) or Sr (99.99\%) were mixed, ground and packed into an
alumina tube, which was then sealed in a quartz tube. The mixture was heated
up to 700$^{\circ}$C in 10\,hrs, reacted at this temperature for 48\,hrs, and
finally quenched to room temperature. Single crystals were grown using a
self-flux method similar to that described previously
\cite{Wang-PRB-2011,Wang-PRBR-2011}. Starting materials of Ca or Sr (99.99\%),
Mn (99.9\%), and excess Bi (99.99\%) were mixed in a molar ratio of Sr:Mn:Bi =
1:1:8, and put into an alumina tube before sealing in a quartz tube. The
mixture was heated up to 800$^{\circ}$C in 10\,hrs, held at this temperature
for 5\,hrs, then slowly cooled to 450$^{\circ}$C at a rate of 3$^{\circ}%
$C\,hr$^{-1}$. The excess Bi flux was decanted at this temperature in a
centrifuge. These materials are reactive in air so handling was carried out in
an inert gas atmosphere as far as possible.

\begin{table}[t]
\centering
\renewcommand{\arraystretch}{1.5}
\begin{tabular}
[c]{ccc}\hline\hline
& \hspace{10pt}CaMnBi$_{2}$ & \hspace{10pt}SrMnBi$_{2}$\\\hline
Ca/Sr & \hspace{10pt}25.1(0.3) & \hspace{10pt}25.6(0.3)\\
Mn & \hspace{10pt}25.4(0.2) & \hspace{10pt}26.6(0.3)\\
Bi & \hspace{10pt}49.5(0.4) & \hspace{10pt}47.8(0.5)\\
& \hspace{10pt}100\% & \hspace{10pt}100\%\\\hline\hline
\end{tabular}
\caption{Electron-probe microanalysis (EPMA) of the composition of
CaMnBi$_{2}$ and SrMnBi$_{2}$ single crystals. The results are given in atom
\%, and are averages over 10 (CaMnBi$_{2}$) or 12 (SrMnBi$_{2}$) points on the
crystal surface. The standard deviations, given in parentheses, show the
compositional spread and indicate the experimental error.}%
\label{table1}%
\end{table}

The crystals were confirmed as single phase by room temperature X-ray
diffraction measurements on powdered crystals.
To check their composition, electron-probe microanalysis (EPMA) was performed
at 10--12 points on the clean surface of one crystal of each type. The
measured cation ratios (in atom \%) are given in Table \ref{table1}. Both
crystals are very close to the ideal stoichiometry, although the data suggest
a small ($\sim$2\%) Bi deficiency in SrMnBi$_{2}$. The analysis also revealed
oxygen on the surface which most likely formed during brief exposure to air.

Magnetic susceptibility measurements were performed with a Superconducting
Quantum Interference Device (SQUID) magnetometer. The susceptibility was
measured under zero-field-cooled (ZFC) and field-cooled (FC) conditions, with
the measuring field applied either parallel or perpendicular to the $c$-axis.
Measurements of the in-plane resistivity ($\rho_{ab}$) were made by the
standard 4-probe method. Neutron time-of-flight diffraction data were
collected on 3\thinspace g powder samples of CaMnBi$_{2}$ and SrMnBi$_{2}$,
and on a $1\times2\times2$\thinspace mm single crystal of SrMnBi$_{2}$. The
measurements were performed on the WISH diffractometer \cite{Chapon-NN-2011}
at the ISIS Facility of the Rutherford Appleton Laboratory (UK).

First principles calculations were performed using the WIEN2k
package\cite{Wien2k}, including the Generalized Gradient Correction to the DFT
and spin-orbit interaction, with $k$-point meshes up to $58\times58\times11$. The magnetic
field in all calculations was assumed to be parallel to $c.$

\section{Results and Analysis}

\begin{figure*}
\includegraphics[clip,angle=0,width=0.4\textwidth]{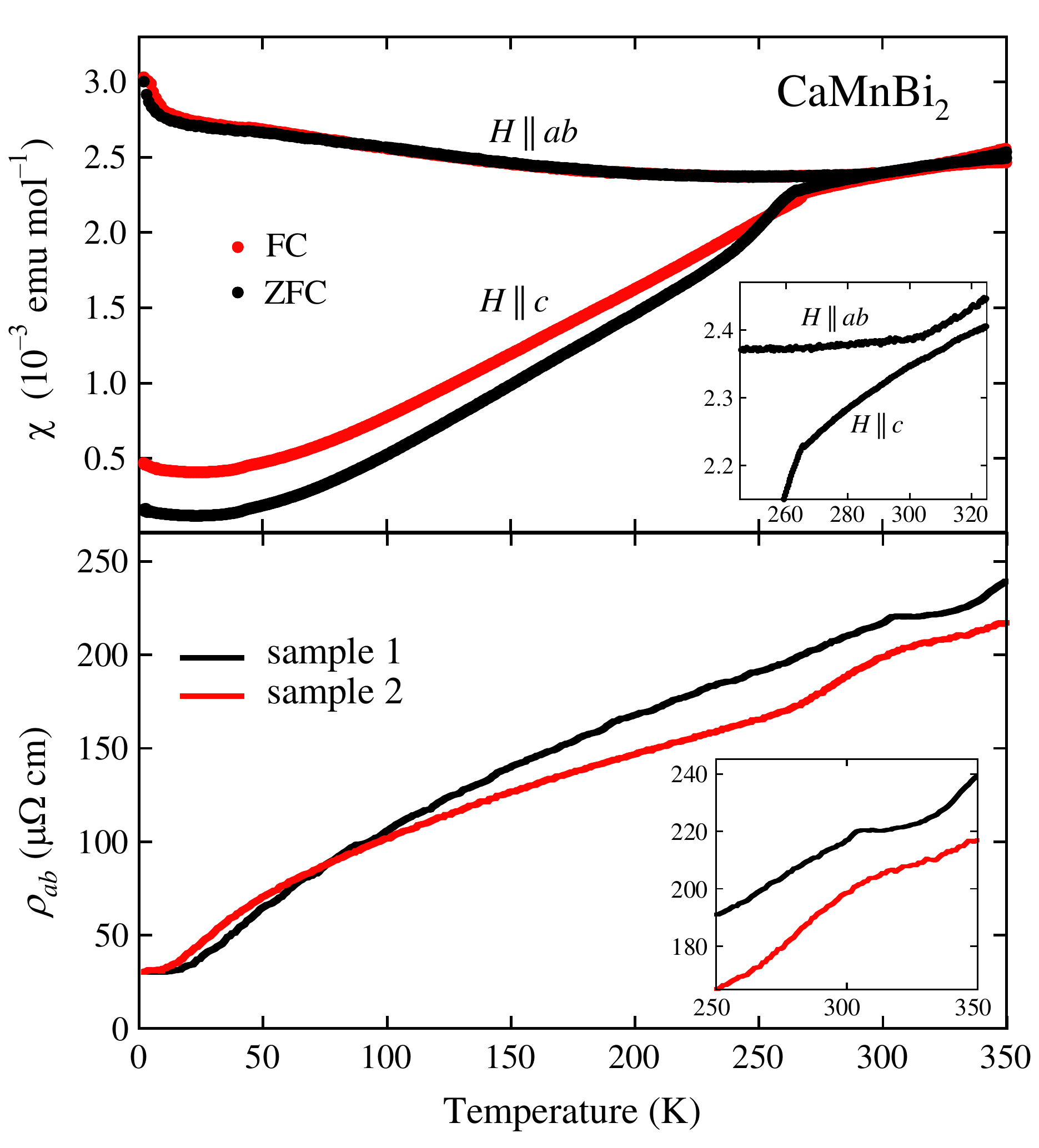}
\includegraphics[clip,angle=0,width=0.4\textwidth]{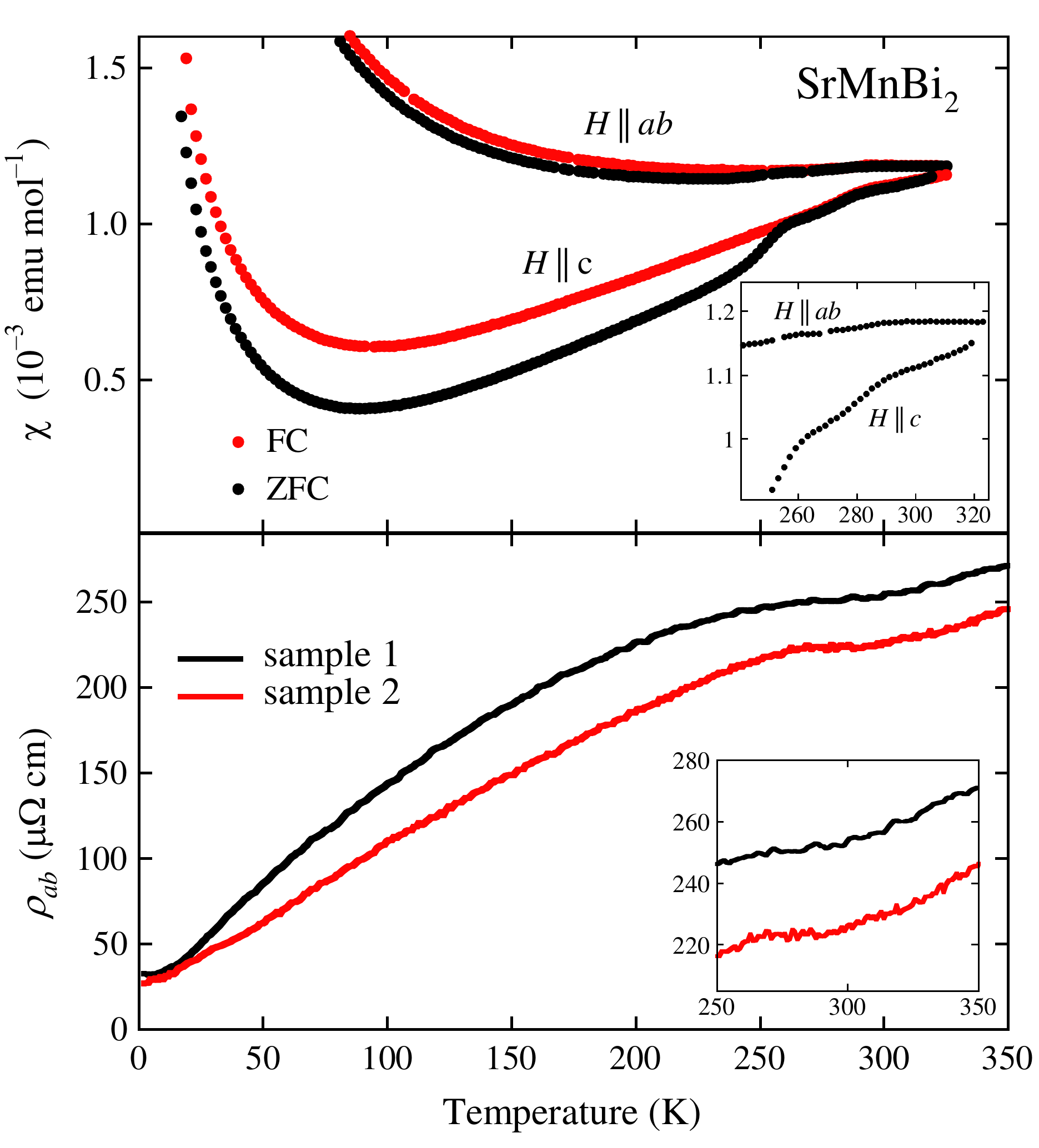}\caption{(Color
online) Temperature dependence of the susceptibility and resistivity of
CaMnBi$_{2}$ and SrMnBi$_{2}$. The susceptibility (upper panels) was measured
under zero-field-cooled (ZFC) and field-cooled (FC) conditions in a field of
$H = 10$\,kOe ($\mu_{0} H = 1$\,Tesla) applied parallel to the $ab$ plane ($H
\parallel ab$) and along the $c$ axis ($H \parallel c$). The susceptibility
insets show the ZFC data on an expanded scale in the vicinity of the magnetic
anomalies at $T_{1} \approx300$\,K and $T_{2} \approx260$\,K. The in-plane
resistivity (lower panels) was measured on two different samples for each
material. The resistivity insets show the data on an expanded scale in the
vicinity of the magnetic anomalies.}%
\label{rhochi}%
\end{figure*}

Figure \ref{rhochi} (upper panels) shows the temperature dependence of the
anisotropic magnetic susceptibility of CaMnBi$_{2}$ and SrMnBi$_{2}$ measured
with a field of $H = 10$\,kOe ($\mu_{0} H = 1$\,Tesla) applied parallel to the
$ab$ plane ($H \parallel ab$) and along the $c$ axis ($H \parallel c$). Both
materials have similar susceptibility curves. There are anomalies at two
temperatures $T_{1}$ and $T_{2}$, where $T_{1} = 305 \pm5$\,K (CaMnBi$_{2}$)
and $295 \pm5$\,K (SrMnBi$_{2}$), and $T_{2} = 265 \pm5$\,K (CaMnBi$_{2}$) and
$260 \pm5$\,K (SrMnBi$_{2}$). These can be viewed on an expanded scale in the
susceptibility insets. From our neutron diffraction data (see below) we
identify $T_{1}$ with the onset of antiferromagnetic order at the N\'{e}el
temperature $T_{\mathrm{N}}$. Below $T_{1} = T_{\mathrm{N}}$ the
susceptibility is strongly anisotropic, with $\chi_{c} < \chi_{ab}$. Below
$T_{2}$ there is a prominent splitting between ZFC and FC measurements for $H
\parallel c$. The $T_{1}$ and $T_{2}$ anomalies have both been reported
previously for SrMnBi$_{2}$ (Ref.~\onlinecite{Wang-PRB-2011}), but only the
$T_{2}$ anomaly has been reported before now for CaMnBi$_{2}$. In common with
previous data,\cite{Park-PRL-2011,Wang-PRB-2011} the susceptibility of
SrMnBi$_{2}$ shows a strong Curie contribution at low temperatures. This
indicates the presence of a small amount of Mn-containing paramagnetic
impurity which might be related to the slight Bi deficiency indicated by EPMA
(Table~\ref{table1}).

Measurements of the in-plane resistivity ($\rho_{ab}$) of two different
samples of each material are presented in Fig.~\ref{rhochi} (lower panels).
The data for samples 1 and 2 of each material are broadly consistent with one
another and with previous
studies,\cite{Park-PRL-2011,Wang-PRB-2011,Wang-PRBR-2011,Wang-PRBR-2012,He-APL-2012,Wang-APL-2012}%
, but there are differences in some details. Firstly, our measurements, which
extend above 300\,K, reveal a bump at $T_{1} = 305$\,K for CaMnBi$_{2}$ which
is not present in the data for SrMnBi$_{2}$. The anomaly is particularly sharp
for sample 1, but is present for both samples of CaMnBi$_{2}$. The resistivity
of the SrMnBi$_{2}$ samples is smoother around room temperature with a small
positive curvature that contrasts with the bump in the CaMnBi$_{2}$ data.
Second, there also appear to be features near $T_{2}$ in the resistivity of
sample 2 of both materials. However, the curves for samples 1 and 2 are not
consistent in this temperature range, and there are no corresponding features
near $T_{2}$ in previous data for CaMnBi$_{2}$ or SrMnBi$_{2}$. We assume,
therefore, that these features are not intrinsic, and speculate that they
could be effects due to the contacts. Finally, there have been reports of an
anomaly in the resistivity of CaMnBi$_{2}$ between 40 and
50\,K.\cite{Wang-PRBR-2012,He-APL-2012,Wang-APL-2012} However, no
corresponding anomalies in the heat capacity have been reported, and we do not
observe such an anomaly in our data.

\begin{figure}
\includegraphics[clip,angle=0,width=0.45\textwidth]{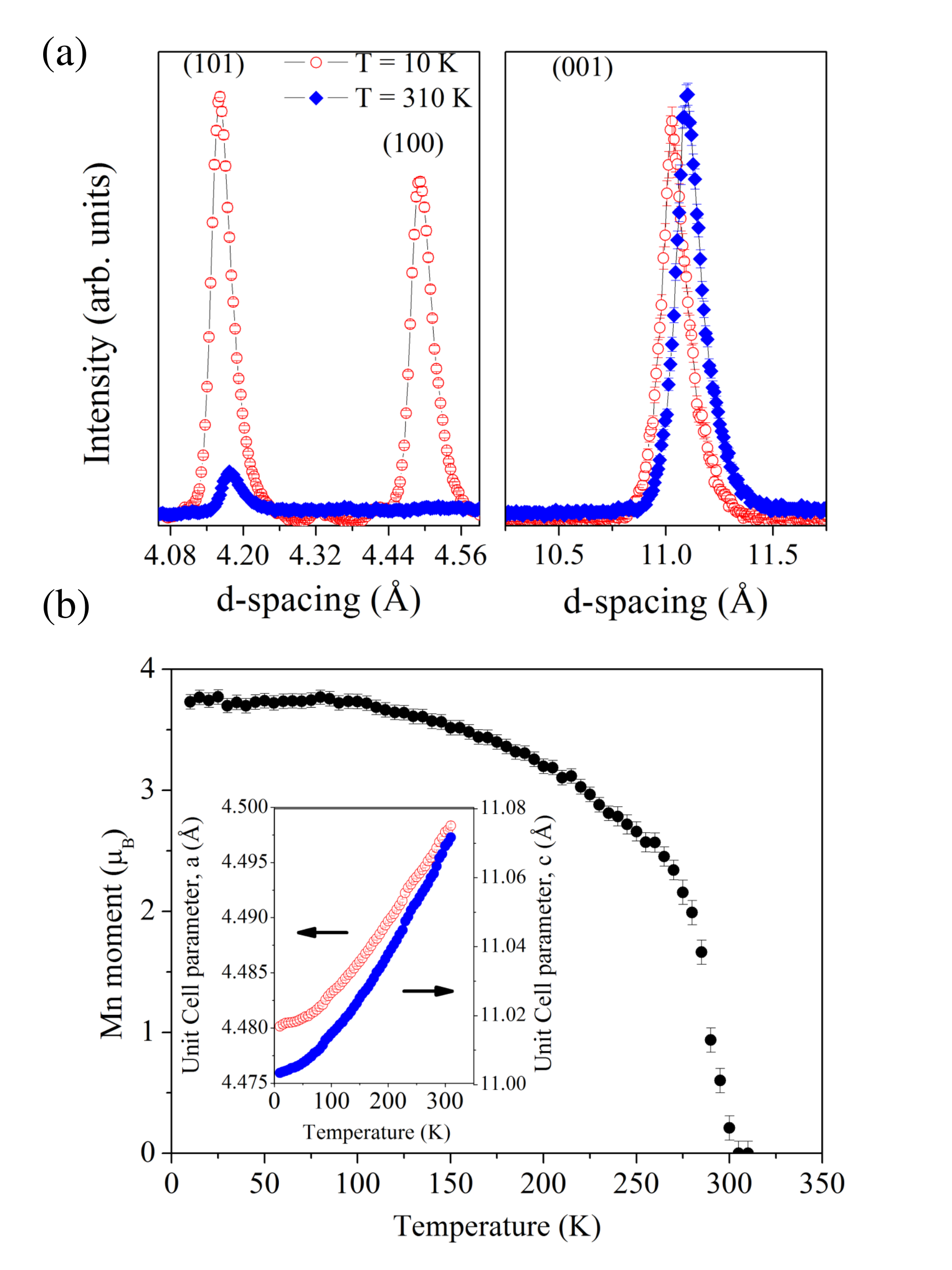}\caption{(Color
online) Neutron powder diffraction data for CaMnBi$_{2}$. (a) Left panel:
$(100)$ and $(101)$ magnetic Bragg peaks at $10$\,K and $310$\,K. Right panel:
$(001)$ structural Bragg peak at $10$\,K and $310$\,K showing absence of a
magnetic contribution. (b) Temperature dependence of the refined magnetic
moment. The insert shows the temperature dependence of the $a$ and $c$ lattice
parameters. }%
\label{fig1}%
\end{figure}

\begin{figure}
\includegraphics[width=0.5\textwidth]{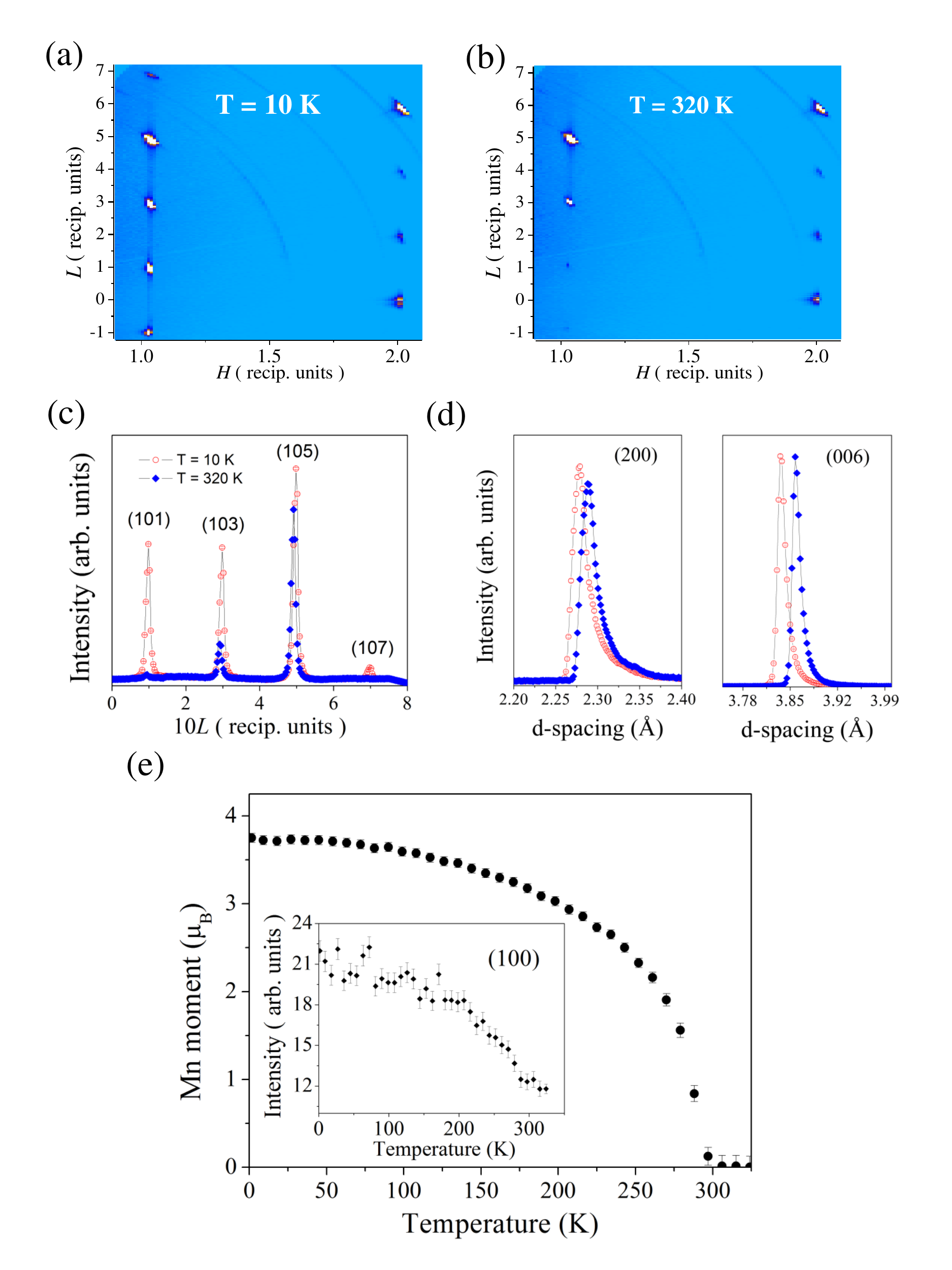}\caption{(Color online) Neutron
single-crystal diffraction data for SrMnBi$_{2}$. Diffraction peaks in the
$(H,0,L)$ scattering plane at (a) $T=10$\,K and (b) $T = 320$\,K. (c) Line
scan along $(1,0,L)$ at $10$\,K and $320$\,K, showing additional magnetic
intensity in the magnetically ordered phase. (d) $(200)$ and $(006)$
reflections showing absence of a magnetic contribution to these peaks. (e)
Temperature dependence of the refined magnetic moment. The insert shows the
temperature dependence of the weak $(100)$ magnetic reflection.}%
\label{data}%
\end{figure}

\begin{figure}
\includegraphics[clip,width=0.45\textwidth]{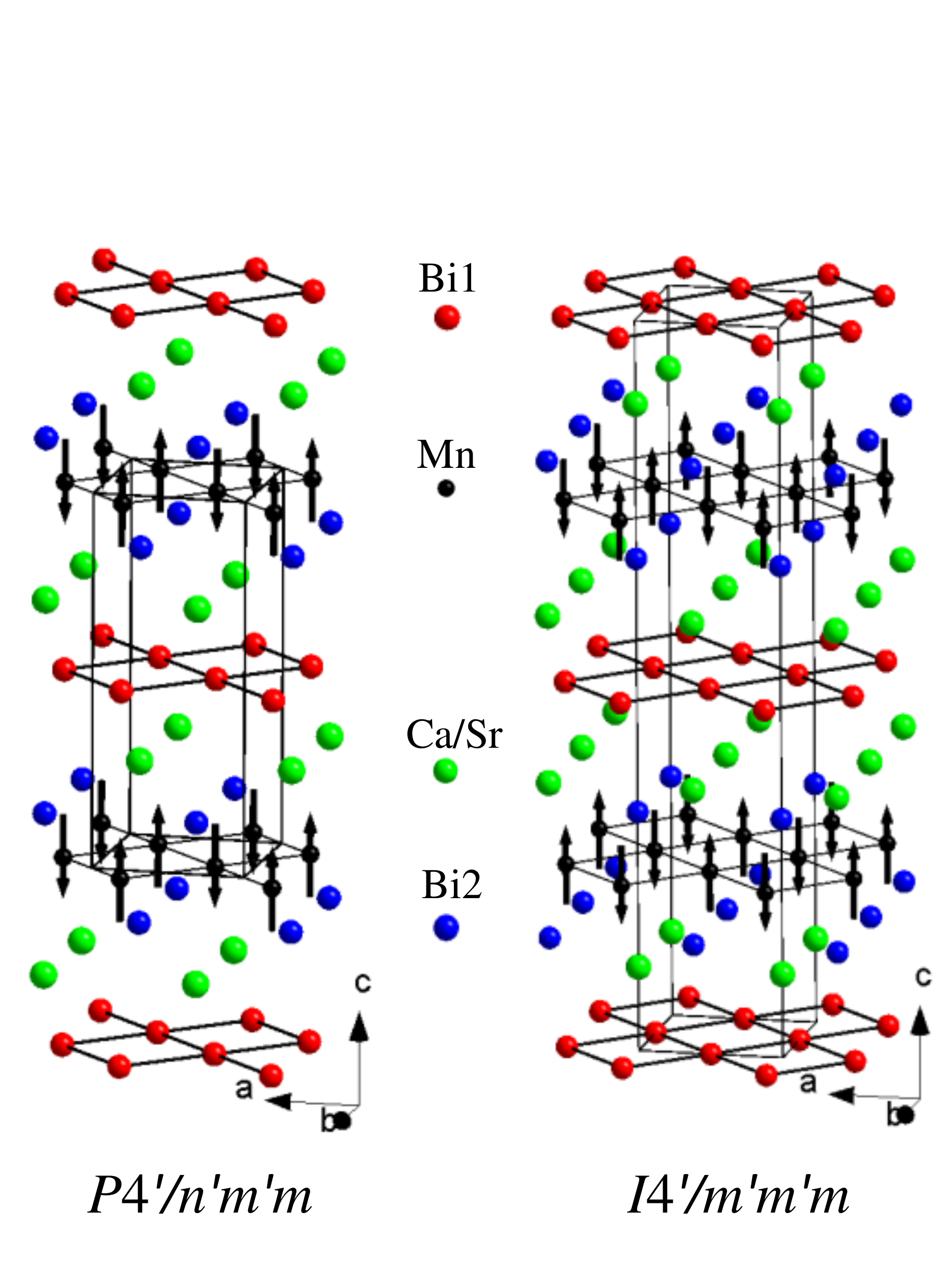}\caption{(Color online)
Magnetic structures of CaMnBi$_{2}$ (left) and SrMnBi$_{2}$ (right). }%
\label{model}%
\end{figure}

The neutron powder diffraction pattern of CaMnBi$_{2}$ collected in the
paramagnetic phase at $T=310$\,K was fitted with the structural model proposed
by Brechtel \textit{et al.} \cite{Brechtel-ZNatur-1980}. The model implies
tetragonal $P4/nmm$ symmetry with Mn occupying 2$a$ Wyckoff sites at
$3/4,1/4,0$ and $1/4,3/4,0$ in the unit cell. The room temperature lattice
parameters were refined as $a=4.4978(1)$\,{\AA } and $c = 11.0692(6)$\,{\AA },
where the numbers in parentheses are fitting errors (one standard deviation).
Below $T_{\mathrm{N}}$ additional scattering appears as illustrated in
Fig.~\ref{fig1}(a), left panel, revealing the onset of magnetic ordering with
a $\mathbf{k} = 0$ propagation vector. The lack of a magnetic contribution to
the $(001)$ reflection, Fig.~\ref{fig1}(a), right panel, implies that the
magnetic moments point along the $c$-axis, and the strong magnetic intensity
at the nuclear-forbidden $(100)$ reflection points to an antiferromagnetic
coupling between the two symmetry-related Mn sites.

These observations unambiguously determine the magnetic structure shown in
Fig.~\ref{model} (left). The model has antiferromagnetic in-plane and
ferromagnetic out-of-plane coupling between the nearest neighbors, and is
described by the $P4^{\prime}/n^{\prime}m^{\prime}m$ magnetic space group. The
refined value of the moment size is $3.73(5)$\,$\mu_{\mathrm{B}}$ at
$T=10$\,K, and the temperature dependence of the moment is shown in
Fig.~\ref{fig1}(b). The AFM ordering temperature in our sample, $T_{\mathrm{N}%
} = 300 \pm5$\,K, is consistent with the susceptibility and resistivity
anomalies at $T_{1}$ (Fig.~\ref{rhochi}), identifying $T_{1}$ with the
N\'{e}el temperature $T_{\mathrm{N}}$. $T_{\mathrm{N}}$ of CaMnBi$_{2}$ is therefore $\sim
$30\,K higher than previously reported based only on susceptibility
data\cite{Wang-PRBR-2012,He-APL-2012}.

Initial neutron powder diffraction measurements on SrMnBi$_{2}$ revealed a
number of candidate magnetic peaks. Subsequently, single crystal neutron
diffraction data on SrMnBi$_{2}$ were collected in two scattering geometries
to access the $(H,K,0)$ and $(H,0,L)$ scattering planes. Room temperature
lattice parameters refined in the $I4/mmm$ space group were found to be
$a=4.5771(2)$\,{\AA } and $c = 23.14069(5)$\,{\AA }. Regions of the $(H,0,L)$
plane measured at temperatures of 10\,K and 320\,K are shown in
Figs.~\ref{data}(a) and (b), respectively. A strong magnetic contribution to
the nuclear reflections along the $(1,0,L)$ line was observed at $T=10$\,K,
see Fig.~\ref{data}(c). The magnetic intensity is resolution-limited and
decreases with increasing scattering vector as expected due to the magnetic
form factor. Based on this observation, a magnetic ordering of the Mn
sublattice with the propagation vector $\mathbf{k} =0$ can be concluded.
Inspection of the $(H00)$, $(0K0)$ and $(00L)$ reflections did not reveal any
magnetic contributions, Fig.~\ref{data}(d). The slightly larger intensity of
the $(200)$ reflection at 10\,K is of structural origin (uncorrelated atomic
displacements) since the same thermal effect is observed for the $(400)$
reflection (not shown).

To obtain a model for the magnetic structure of SrMnBi$_{2}$ we adopted a
symmetry-based approach, analysing the magnetic reflection conditions for the
possible magnetic space groups. The parent symmetry was assumed to be
$I4/mmm$, as determined by Cordier and Sch\"{a}fer \cite{Cordier-ZNatur-1977}.
The magnetic space groups associated with $\Gamma$-point ($\mathbf{k} = 0$)
were generated by the ISOTROPY software \cite{isotropy,Campbell-JAC-2006} for
the irreducible representations entering the pseudovector reducible
representation on the 4$d$ Mn position. Then, the extinction rules for the
magnetic space groups tabulated in the Bilbao Crystallographic Server (MAGNEXT
\cite{MAGNEXT}) for non-polarized neutron diffraction were applied, resulting
in the unambiguous choice of $I4^{\prime}/m^{\prime}m^{\prime}m$ as the
appropriate magnetic symmetry for SrMnBi$_{2}$. This space group is associated
with the one-dimensional $\Gamma_{2}^{-}$ irreducible representation and
implies an antiferromagnetic arrangement for both the in-plane and
out-of-plane nearest neighbours with the spin direction along the $c$-axis, as
shown in Fig.~\ref{model} (right). The refined value of the magnetic moment at
$T=10$\,K is $3.75(5)\,\mu_{\mathrm{B}}$, and the temperature dependence of
the moment is shown in Fig.~\ref{data}(e). The AFM ordering temperature
$T_{\mathrm{N}} = 295 \pm5$\,K is consistent with previous reports
\cite{Park-PRL-2011,Wang-PRB-2011} and with the value of $T_{1}$ from the
magnetic susceptibility. The saturated moment is the same to within
experimental error as we find in CaMnBi$_{2}$. Thus, the main difference
between the magnetic structures of CaMnBi$_{2}$ and SrMnBi$_{2}$ is the sign
of the out-of-plane coupling: ferromagnetic for the former,
antiferromagnetic for the latter.

In addition to the $\mathbf{k} = 0$ magnetic peaks, we observed a very small
$(100)$ reflection in the data for SrMnBi$_{2}$. The $(100)$ is forbidden in
the $I$-centered lattice, and is only observed at $T < T_{\mathrm{N}}$ --- see
insert to Fig.~\ref{data}(e). This observation indicates the existence of a
structural distortion with wave vector $\mathbf{k} = (1,1,1)$. If the magnetic
transition is continuous, as suggested by Fig.~\ref{data}(e), then either (i)
there exists a structural instability unrelated to the magnetic order that
occurs very close to (but not coincident with) $T_{\mathrm{N}}$, or (ii) the
primary $\mathbf{k} = 0$ magnetic order parameter induces (via a trilinear
free energy invariant) a secondary magnetic mode with $\mathbf{k} = (1,1,1)$
due to the existence already in the paramagnetic phase of a structural
distortion also with $\mathbf{k} = (1,1,1)$. Group theoretical analysis shows
that in the latter case the symmetry of the paramagnetic phase would have to
be $P4/nmm$, $Cmcm$ or $Pmmn$. However, we failed to find any direct evidence
that the high-temperature structure is other than $I4/mmm$.

\begin{figure*}
\includegraphics[clip,angle=0,width=0.8\textwidth]{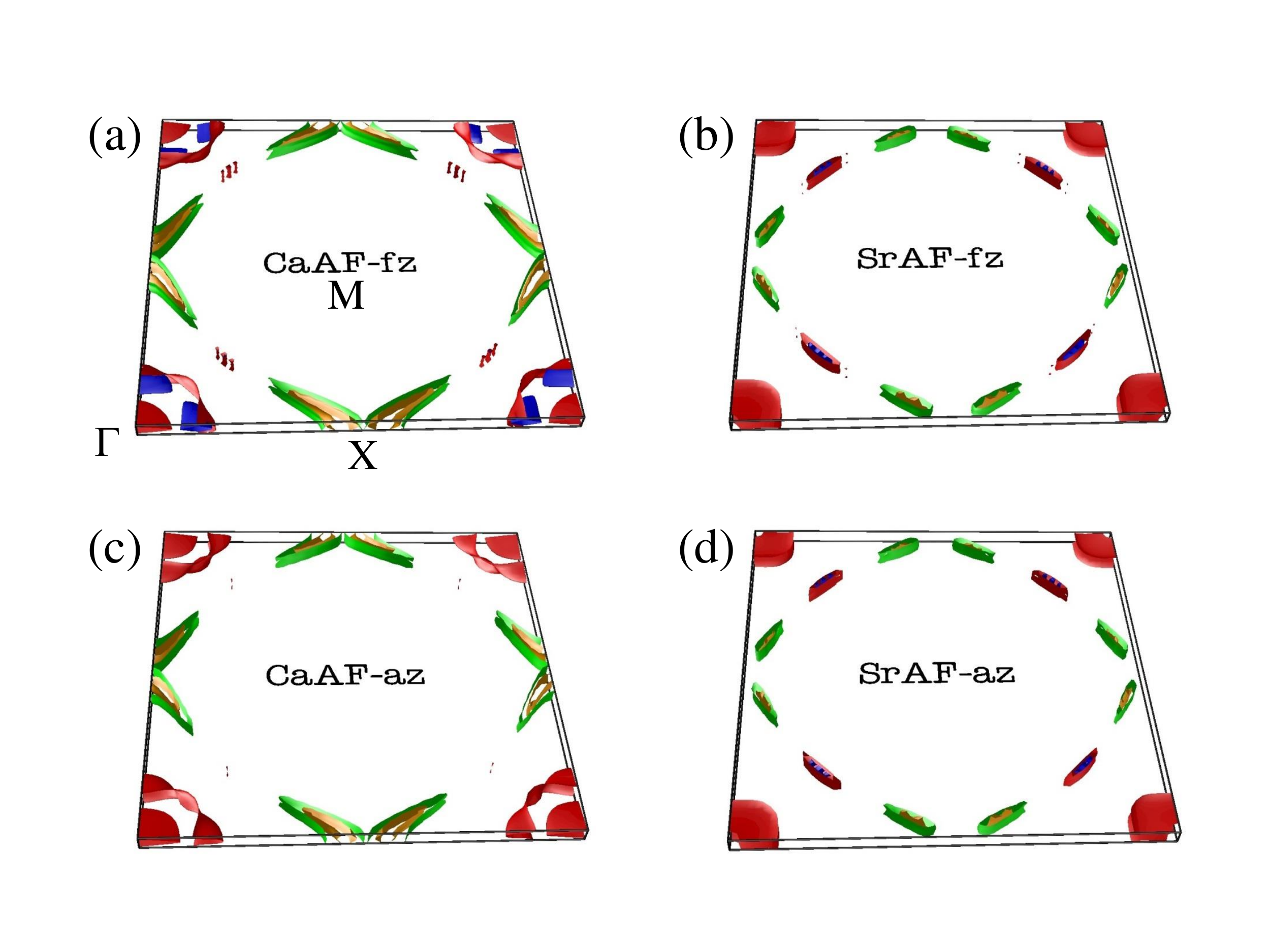}
\caption{(Color online) Calculated Fermi surfaces of CaMnBi$_{2}$ and SrMnBi$_{2}$ with N\'{e}el ordering in the plane and either
ferromagnetic inter-layer stacking, (a) and (b), denoted \textquotedblleft AF-fz", or antiferromagnetic stacking, (c) and (d), denoted \textquotedblleft AF-az". Note that the Fermi surface of SrMnBi$_{2}$ is virtually sensitive to the stacking, but that of CaMnBi$_{2}$ is not.}%
\label{FS}%
\end{figure*}

\begin{figure}
\includegraphics[clip,angle=0,width=0.45\textwidth]{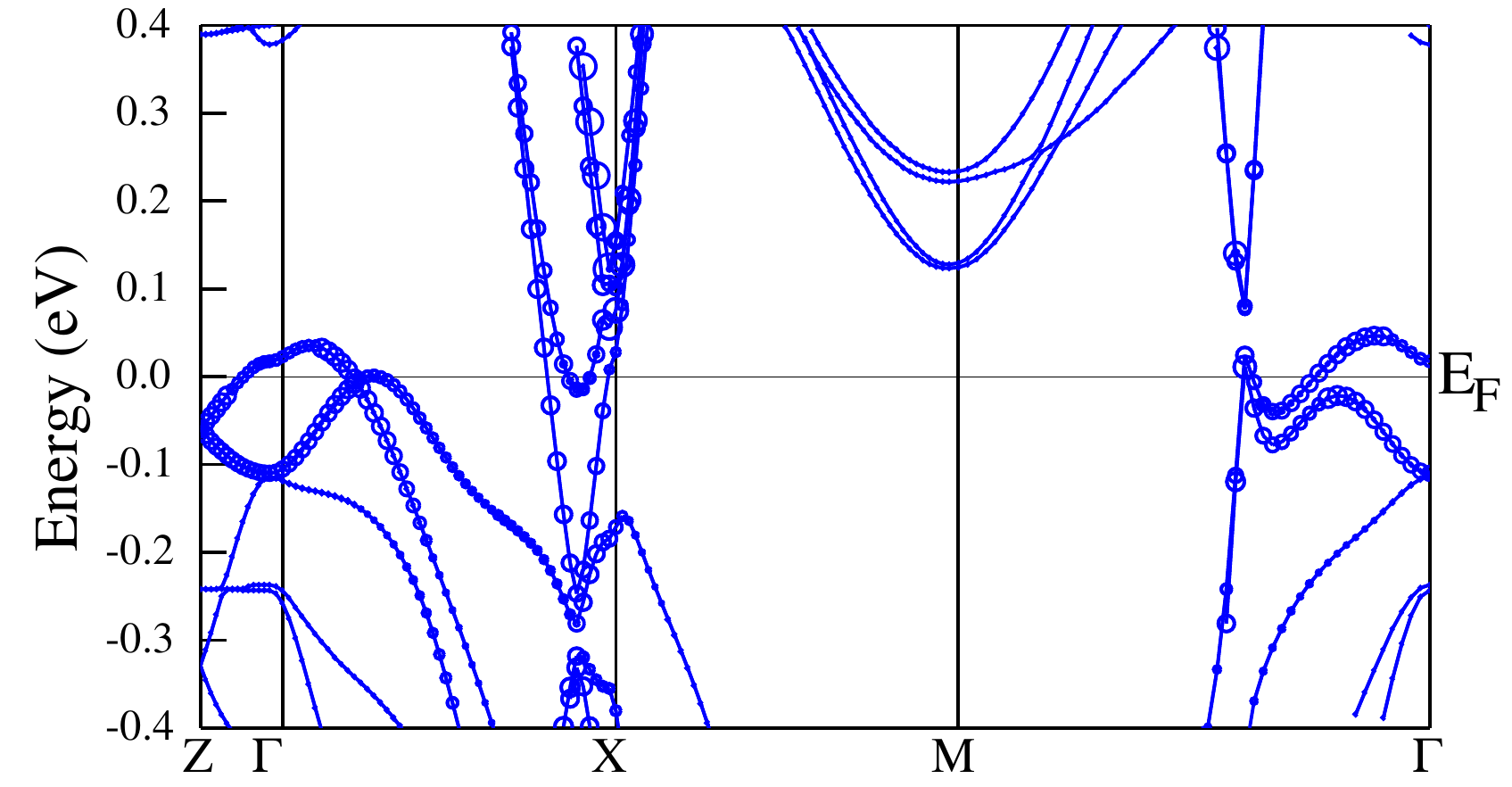}
\caption{(Color online) Band structure of CaMnBi$_{2}$ calculated with ferromagnetic stacking (AF-fz). The size of the circles represents the contribution from
the Bi $6p_{x,y}$ orbitals.}%
\label{bands}%
\end{figure}

Next, we describe the results of the band structure calculations. Figures~\ref{FS}(a) and (b) show the Fermi surfaces of both compounds calculated with antiferromagnetic order in the plane and ferromagnetic
stacking along $c$ (denoted \textquotedblleft AF-fz"). Figures~\ref{FS}(c) and (d) show likewise for antiferromagnetic stacking (\textquotedblleft AF-az"). For the sake of comparison, we used for both compounds a tetragonal cell containing two Mn layers,
corresponding to one (SrMnBi$_{2}$) and two (CaMnBi$_{2}$) of the unit cells depicted in Fig.~\ref{model}.

To illustrate the band structure in the vicinity of the Fermi energy, $E_{\rm F}$, we show in Fig.~\ref{bands}
 the calculated band dispersion of CaMnBi$_{2}$ for energies in the range $E_{\rm F} \pm 0.4$\,eV for
the case of ferromagnetic stacking. The results are consistent with earlier calculations.
\cite{Wang-PRBR-2012,Lee-PRB-2013} (note that the bands in Fig.~\ref{bands} are downfolded
along $k_z$ compared to those in Refs. \onlinecite{Wang-PRBR-2012,Lee-PRB-2013}). The Dirac cones are located between the $\Gamma$ and M points, and are strongly squeezed in the
$(110)$ direction. Moreover, the Dirac points (though not the Dirac bands) are
destroyed by the spin-orbit interaction (a $k_{z}$-dependent gaps opens with magnitude
 varying between 1.4 and 15 meV). At the same time, several other Fermi surface pockets, besides the
 Dirac ones, are also formed by the planar Bi electrons, but these have strong $p_z$ character, as opposed to the
$p_{x,y}$-derived Dirac bands.

To assess the relative importance of the Dirac and non-Dirac bands to the electronic transport
we have calculated for CaMnBi$_{2}$ the band-decomposed plasma frequencies $\omega_p$. In the constant scattering rate approximation the conductivity is proportional to $\omega_p^2$. We found that for the Dirac bands
$\omega_{px}=\omega_{py}=2.45$\,eV, and $\omega_{pz}=0.24$\,eV. For the non-Dirac bands these
numbers are 0.42 and 0.30 eV, respectively. Thus, the in-plane transport is dominated by the
Dirac electrons (as opposed to the out-of-plane one). Unfortunately, it is harder to decompose
the plasma frequency for SrMnBi$_{2}$ in a similar way, because some of the Dirac bands have the same band number
as the 3D bands (cf. the colors in Fig. \ref{FS}).



\begin{figure}
\includegraphics[clip,angle=0,width=0.45\textwidth]{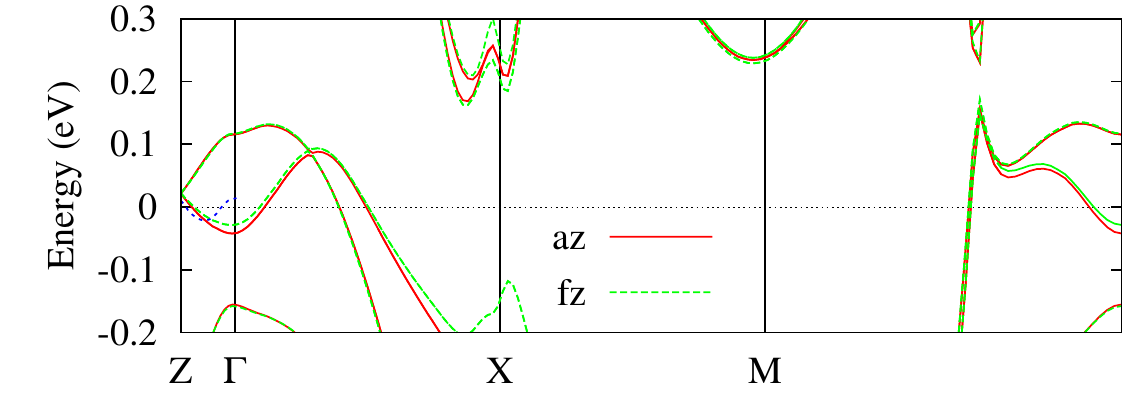}
\includegraphics[clip,angle=0,width=0.45\textwidth]{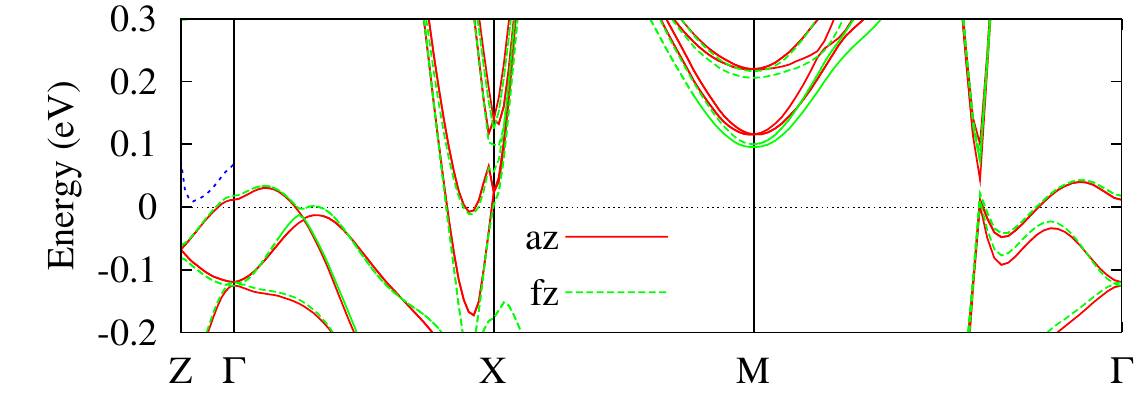}
\caption{(Color online) Effect of ferromagnetic $vs$ antiferromagnetic stacking on the band structure of
SrMnBi$_{2}$ (upper panel) and CaMnBi$_{2}$ (lower panel). The dotted (blue) lines between $\Gamma$ and Z represent the difference between the fz and az bands multiplied by 10 to emphasize the stacking-dependent broadening of the three-dimensional Bi band near $\Gamma$ for CaMnBi$_{2}$ relative to SrMnBi$_{2}$.}%
\label{bands2}%
\end{figure}

Interestingly, switching from ferromagnetic to antiferromagnetic inter-layer stacking has a rather distinct effect on the fermiology of the two compounds.
In SrMnBi$_{2}$, the Fermi surface is virtually insensitive to the type of stacking, whereas in CaMnBi$_{2}$ there is a clear difference in the bands near $E_{\rm F}$ for the two types of stacking. This difference can be seen in the Fermi surfaces, Fig.~{\ref{FS}, and in the stacking-dependent band dispersions shown in Fig.~\ref{bands2}. The strongest effect is seen on the three-dimensional Fermi surface
pocket near $\Gamma$. One can also see that for CaMnBi$_{2}$ the dispersion of this band along $z$ is higher in the fz structure (by $\approx10\%$) than in the az one, whereas for SrMnBi$_{2}$ the same band is very similar for the fz and az structures. This feature could prove important in the energetics, as discussed below.
These results indicate that the bands near $E_{\rm F}$ couple more strongly to the inter-layer magnetic order in CaMnBi$_{2}$ than in SrMnBi$_{2}$, in agreement with the experimental observation of a resistivity anomaly at $T_{\rm N}$ for CaMnBi$_{2},$ but not for SrMnBi$_{2}.$

We have calculated the energy difference between the two different magnetic stackings in both compounds. Independently of the stacking, we found the moment inside the muffin-tin sphere of Mn (radius $1.3 \times 10^{-10}$\,m) to be 4.03 $\mu_{B}$ for SrMnBi$_{2}$ and 3.97 $\mu_{B}$ for CaMnBi$_{2}$. The deviation from experiment is about 7\%, which tells us that fluctuations beyond the mean field are weak (compared for example with the Fe pnictides where the deviation is close to a factor of two). We found that the energy difference per Mn is 1.2\,meV for
CaMnBi$_{2}$, in favor of the ferromagnetic stacking, and 1.5\,meV for SrMnBi$_{2}$, in favor of the antiferromagnetic stacking, both in agreement with experiment. We have also calculated the energy cost of aligning all spins ferromagnetically and found it to be extremely high, on the order of 300\,meV per Mn for both compounds, suggesting that these systems are magnetically extremely two-dimensional, and that the relatively low ordering temperature is due to the logarithmic suppression of the N\'{e}el temperature due to inter-layer fluctuations.

We have established that the coupling of the three-dimensional Bi bands to the Mn magnetic order is stronger in CaMnBi$_{2}$ than SrMnBi$_{2}$. But why is there a reversal of the inter-layer interaction? First, one may think about the Hund's coupling on the planar Bi. Indeed, in the fz configuration the Bi is allowed to acquire a magnetic moment, thus gaining magnetic energy of $M_{\rm Bi}^{2}I_{\rm Bi}/4$, where $I_{\rm Bi}\lesssim 1$\,eV is the Stoner parameter for Bi. Indeed, Bi does acquire a magnetic moment, and according to our calculations it is a factor of 2 larger in CaMnBi$_{2}$ than in SrMnBi$_{2}$ (0.007 $\mu_{\rm B}$ $vs.$ 0.003 $\mu_{\rm B})$, but the corresponding energy gain is less than 20\,$\mu$eV, not enough by far to explain the effect.

We propose instead that the ferromagnetic interaction between the layers in CaMnBi$_{2}$ is similar in nature to double exchange and to ferromagnetism in dilute magnetic semiconductors. The fact that the relevant Bi band becomes some 10\% wider in the ferromagnetically-stacked CaMnBi$_{2}$ indicates better delocalization of the corresponding electrons and therefore a gain in their kinetic energy. The order of magnitude of this effect can be obtained from the number of holes in the $\Gamma$-centered Fermi surface pocket multiplied by the width of the relevant band. This rough estimate gives 3--5\,meV, which is in the right ballpark. The virtual absence of any coupling of the Bi electrons to magnetism in SrMnBi$_{2}$ suggests that the antiferromagnetic inter-layer coupling in SrMnBi$_{2}$ is caused by superexchange. We conclude, therefore, that the inter-layer interaction comes about from competition between the antiferromagnetic standard superxchange and a double-exchange-like itinerant ferromagnetic interaction.

The latter interaction may not be very accurately described by a short-range Heisenberg interaction, but given its small amplitude this is not a bad model. In this case, the minimal model is the square 2+1D model, $\mathcal{H}=\sum_{\rm nn}J_{ab}S_{i}S_{j}+ \sum_{\rm nn}J_{c}S_{i}S_{j}$,
where the former sum is taken over all nearest-neighbor (nn) bonds in the plane and the latter over all such bonds between the planes. From the energy differences between fz and az configurations we can deduce
$J_{c}S^{2}=(E_{\rm fz}-E_{\rm az})/2\sim\pm0.7$ meV (there is one such bond per Mn).
We can also estimate $J_{ab}$ from the calculated energy difference between
a ferromagnetic and an antiferromagnetic in-plane arrangement. We found this
difference to be about 300\,meV for both compounds, so that $J_{ab}
S^{2}=(E_{\rm FM}-E_{\rm AF})/4 \approx 75$\,meV.

There have been numerous studies of such 2+1D models. The Monte-Carlo simulations of Yasuda $et$ $al.$\cite{Yasuda}, consistent with the analytical results of Irkhin and Katanin\cite{IK}, suggest that for the ratio $J_{ab}/|J_{c}|\sim 100$ the transition temperature is $T_{\rm N}\approx 0.7J_{ab}S^2/k_{\rm B}\sim 600$\,K.
This is about twice larger than the experimental number, but is in fact very
consistent with it: first, experimental moments are smaller than the
calculated mean-field ones by about 7\%, which suggests that the exchange energy
scale is suppressed by fluctuations by some 15\%, reducing $T_{\rm N}$ to
$\sim 500$\,K. Second, the superexchange interaction is inversely proportional to the energy cost of moving a Mn electron to another atom while flipping its spin,
 $\Delta_{\uparrow\downarrow}$. This is
routinely underestimated in DFT calculations because of insufficient account
of the on-site Hubbard repulsion. For Mn$^{2+}$ underestimation of 50--100\% is
common. To demonstrate that, we have performed calculations using the LDA+U
formalism. In this formalism, $\Delta_{\uparrow\downarrow}\approx 5I+U_{\rm eff},$
where $U_{\rm eff}=U-J_{\rm H}$. Here, $I\lesssim 1$\,eV is the DFT Stoner factor for Mn,
and $U$ and $J_{\rm H}$ are the Hubbard repulsion and the Hund's rule coupling on Mn. Indeed, we found
that the energy difference between the ferromagnetic and antiferromagnetic states
follows the same formula, $J_{ab}\propto(5$\,eV$+U_{\rm eff})^{-1}$, and for a very reasonable choice of
$U_{\rm eff}=3$\,eV we obtain $T_{\rm N}\approx 350$\,K, in very good agreement with experiment.

Finally, we have calculated the magnetic anisotropy energy (the difference in energy between a spin pointing parallel and perpendicular to the layers) for CaMnBi$_{2}$
(not including a $U),$ and found it to be $K\approx0.7$ meV per Mn, with the easy
direction the $c$ axis, in agreement with experiment. This value suggests a
spin-flop transition at a field $B_{\rm SF}\approx 2\sqrt{K(E_{\rm FM}-E_{\rm AF})}%
/(g\mu_{\rm B}S)\sim 100$--125\,T.

\section{Conclusions}


The central experimental results of this work are, (i) that the Dirac
materials $A$MnBi$_{2}$ with $A$ = Sr and Ca have N\'{e}el-type in-plane AFM
order (ii) that the MnBi$_{4}$ layers are coupled ferromagnetically in
CaMnBi$_{2}$ but antiferromagnetically in SrMnBi$_{2}$, and (iii) that the
resistivity of CaMnBi$_{2}$ (but not SrMnBi$_{2}$) has an anomaly at
$T_{\mathbf{N}}$. The latter is consistent at the mean-field level with the different inter-layer magnetic couplings. This study also shows conclusively that the AFM ordering
transition correlates with the $T_{1}$ anomalies observed in the
susceptibility.

The opposite inter-layer magnetic coupling in SrMnBi$_{2}$ and
CaMnBi$_{2}$ is fully reproduced in the first principles calculations, and its origin
is suggested to be a competition between antiferromagnetic superexchange and a ferromagnetic
double-exchange-like interaction, the former winning in SrMnBi$_{2}$ and the
latter in CaMnBi$_{2}.$ The ferromagnetic component, itinerant in origin, is
due to a 3D band generated by the square-planar Bi electrons, whose mobility
appears noticeably higher in the ferromagnetically-stacked CaMnBi$_{2}$ than
in the antiferromagnetically-stacked SrMnBi$_{2}$. Our calculations show that
the Dirac fermions dominate the in-plane electrical transport in both materials, but
 magnetism couples largely to non-Dirac-like Bi electrons, consistent with
the relatively small size of the resistivity anomaly observed at $T_{\mathbf{N}}$ in CaMnBi$_{2}$.  It would be of interest to measure the inter-layer transport, which should show a larger effect at $T_{\mathbf{N}}$.

The question of what causes the anomaly at $T_{2}\approx260$\thinspace K
corresponding to the FC--ZFC splitting in the susceptibility remains open. No
heat capacity anomalies have been reported at $T_{2}$, and we could find no
evidence for any magnetic or structural phase changes below $T_{\mathrm{N}}$
to within the sensitivity of our diffraction measurements --- see, for
example, the insert to Fig.~\ref{fig1}(b). Therefore, if these anomalies are
the result of spin reorientations or structural distortions then the changes
to the magnetic or lattice symmetry are very subtle.  The FC--ZFC
splitting at $T_{2}$ is suggestive of domain formation or disorder which might
result from inter-layer stacking faults frozen in at $T_{2}$. We have observed
that there is additional diffuse scattering at 10\thinspace K compared with
300\thinspace K in the form of a rod of scattering along the $(1,0,L)$ line in
reciprocal space --- compare Figs.~\ref{data}(a) and (b). This form of diffuse
scattering is consistent with the existence of stacking faults along the $c$ axis.

\begin{acknowledgments}
We thank Prof. K. Yamaura of the National Institute for Materials Science, Tsukuba, Japan, for performing the EPMA measurements. This work was supported by the U.K. Engineering and Physical Sciences Research
Council. Work in Beijing was supported by the 973 project of the Ministry of
Science and Technology of China (No. 2011CB921701) and the National Natural
Science Foundation of China (No. 11274367). I.I.M. acknowledges funding from
the Office of Naval Research (ONR) through the Naval Research Laboratory's
Basic Research Program.
\end{acknowledgments}


\end{document}